\documentclass[11pt]{article}

\usepackage{amsfonts}
\usepackage{amsthm}
\usepackage{amsmath}
\usepackage[margin=1.4in]{geometry}

\newtheorem{theorem}{Theorem}
\newtheorem{lemma}{Lemma}
\newtheorem{definition}{Definition}

\begin{document}

\title{On the Recognition of Strong-Robinsonian Incomplete Matrices}
\author{Julio Aracena\footnote{CI$^2$MA and Departamento de Ingenier\'ia Matem\'atica, Facultad de Ciencias F\'isicas y Matem\'aticas, Universidad de Concepci\'on, Chile. (\emph{jaracena@ing-mat.udec.cl}). J. Aracena was
partially supported by ANID-Chile through the project {\sc Centro de Modelamiento Matem\'atico} (AFB170001) of the PIA Program:``Concurso Apoyo a Centros Cient\'ificos y Tecnol\'ogicos de Excelencia con Financiamiento Basal'', and by ANID-Chile through Fondecyt project 1151265.
} 
\and Christopher Thraves Caro\footnote{(Corresponding Author) 
Departamento de Ingenier\'ia Matem\'atica,
Facultad de Ciencias F\'isicas y Matem\'aticas,
Univiversidad de Concepci\'on, Chile.
Adress: Av. Esteban Iturra s/n,
Casilla 160-C, Concepci\'on, 
Chile.
Tel.: +56-41-2203129.
Email: \emph{cthraves@ing-mat.udec.cl}.}
}
\date{}

\maketitle


\begin{abstract}
A matrix is \emph{incomplete} when some of its entries are missing. 
A \emph{Robinson} incomplete symmetric matrix is an incomplete symmetric matrix whose non-missing entries do not decrease along rows and columns when moving toward the diagonal. A \emph{Strong-Robinson} incomplete symmetric matrix is an incomplete symmetric matrix $A$ such that $a_{k,l} \geq a_{i,j}$ if $a_{i,j}$ and $a_{k,l}$ are two non-missing entries of $A$ and $i\leq k \leq l \leq j$.
On the other hand, an incomplete symmetric matrix is \emph{Strong-Robinsonian} if there is a simultaneous reordering of its rows and columns that produces a Strong-Robinson matrix.  
In this document, we first show that there is an incomplete Robinson matrix which is not Strong-Robinsonian. 
Therefore, these two definitions are not equivalent.
Secondly, we study the recognition problem for Strong-Robinsonian incomplete matrices.
It is known that recognition of incomplete Robinsonian matrices is NP-Complete. 
We show that the recognition of incomplete Strong-Robinsonian matrices is also NP-Complete.
However, we show that recognition of Strong-Robinsonian matrices can be parametrized with respect to the number of missing entries. Indeed, we present an $O(|w|^bn^2)$ recognition algorithm for Strong-Robinsonian matrices, where $b$ is the number of missing entries, $n$ is the size of the matrix, and $|w|$ is the number of different values in the matrix. 
\end{abstract}
{\bf Keywords:} Robinsonian matrices, strong-Robinsonian matrices, incomplete matrices, matrix completion, recognition problem.  

\section{Introduction}
A symmetric matrix is \emph{Robinson} if its entries
do not decrease along rows and columns when moving toward the main
diagonal. A symmetric matrix is \emph{Robinsonian}
if there is a simultaneous reordering of its rows and columns such that 
it results in a Robinson matrix. 
Robinsonian matrices were defined by W. S. Robinson in  
\cite{robinson_1951} in a study on how to order chronologically 
archaeological deposits. The \emph{Seriation} problem introduced in the 
same work is to decide whether the \emph{similarity} matrix of 
a data set is Robinsonian and write it as a Robinson 
matrix if possible.

Robinsonian matrices have applications in many different contexts such as archaeology \cite{petrie1899sequences}, 
data visualization \cite{brusco2006branch}, exploratory analysis 
\cite{hubert2001combinatorial},
bioinformatics \cite{tien2008methods}, and machine learning 
\cite{ding2004linearized}. Liiv in 
\cite{liiv2010seriation}
 surveyed the 
seriation problem, matrix reordering and its applications. 
Also, Laurent and Seminaroti in   
\cite{laurent2015quadratic}, using Robinsonian matrices, gave
a new class of instances for the Quadratic Assignment Problem which is
solvable in polynomial time.

It is natural to consider the seriation problem with incomplete data due
to errors in the data set \cite{fortin2019clustering,fortin2017robinsonian}. In this document, we extend the definition 
of Robinson and Robinsonian matrices to incomplete symmetric matrices.
A symmetric matrix is said to be \emph{incomplete} if some of its
entries are missing. When considering incomplete symmetric matrices, 
two possible extensions for Robinson matrices appear. The first 
extension says that
an incomplete symmetric matrix is \emph{Robinson} if its non-missing
entries do not decrease along rows and columns when moving toward the 
diagonal. 
But with this definition, for a given entry there may be strictly 
smaller entries that are closer to the main diagonal. This phenomena 
appears due to the missing entries. 
Hence, we present a second extension for Robinson matrices in which this
phenomena is avoided. 
An incomplete symmetric matrix $A$ is \emph{Strong-Robinson} if $a_{k,l}
\geq a_{i,j}$ whenever $a_{i,j}$ and $a_{k,l}$ are two non-missing
entries of $A$ and $i\leq k \leq l \leq j$.
These two definitions coincide when we consider \emph{complete} 
symmetric matrices, and they actually coincide in Robinson matrices. 

From this point, it is natural to define incomplete Robinsonian matrices
and incomplete Strong-Robinsonian matrices as symmetric matrices that
admit a simultaneous reordering of their rows and columns such that 
we obtain an incomplete Robinson matrix or an incomplete Strong-Robinson
matrix, respectively. 

Recognition of incomplete Robinsonian matrices was shown to be
NP-Complete in \cite{cygan2015sitting}. 
Furthermore, they gave a lower bound on the complexity of this problem 
by showing that there is no algorithm running in sub-exponential time 
that recognizes incomplete Robinsonian matrices. 
In this document, we first
understand that these two definitions are actually different and then 
we study the recognition problem for incomplete Strong-Robinsonian
matrices showing that this problem can be solved in polynomial time if 
the number of missing entries is considered to be a 
constant parameter.  

\section{Definitions}\label{sec:definitions}
In this document, we consider symmetric $n\times n$ matrices typically 
denoted by $A$. We use $a_{i,j}$ to denote $A$'s entry in row $i$ and 
column $j$. We say that a matrix is \emph{incomplete} if one or more of 
its entries are not defined, or missing. We use $a_{i,j}=*$ to denote 
that $a_{i,j}$ is a missing entry of $A$. We use $w(A)$ to denote the 
set of different (numerical) values  in $A$ (if $A$ is an 
incomplete matrix the symbol $*$ does not belong to $w(A)$).  
Hence, $|w(A)|$ 
denotes the amount of different numerical values in $A$. As an example, 
if $A$ is a $0-1$ matrix, then $w(A)=\{0,1\}$ and $|w(A)|=2$, even if 
$A$ is incomplete. 
We use $b(A)$ to denote the number of missing entries in $A$. When 
contextually clear, we simply use $b$. 

Now we introduce Robinson and Strong-Robinson incomplete matrices. 
\begin{definition}\label{def:RobinsonIncomplete}
Let $A$ be a symmetric $n \times n$ incomplete matrix. 
We say that $A$ is \emph{Robinson} if:
\begin{itemize}
    \item $a_{i,j} \leq a_{i,k}$ for all $i\leq k \leq j$, 
    such that $a_{i,j} \neq * \neq a_{i,k}$, and
    \item $a_{i,j} \leq a_{l,j}$ for all $i\leq l \leq j$, 
    such that $a_{i,j} \neq * \neq a_{l,j}$.
\end{itemize}  
\end{definition}
\begin{definition}\label{StrongRObinsonIncomplete}
Let $A$ be a symmetric $n \times n$ incomplete matrix. 
We say that $A$ is \emph{Strong-Robinson} if:
\[
a_{i,j} \leq a_{k,l} \mbox{ for all } i\leq k \leq l \leq j 
\mbox{ such that } a_{i,j} \neq * \neq a_{k,l}. 
\]
\end{definition}

Event though, these two definitions look similar, they are not 
equivalent. Nevertheless, if we 
consider \emph{complete} matrices, i. e., matrices with 
no missing entry, these two definitions become equivalent, and they can 
be expressed as follows: 
\[
\mbox{for every } a_{i,j} \mbox{ such that } 
1\leq i < j \leq n: \,\, 
a_{i,j} \leq \min\{a_{i,j-1},a_{i+1,j}\}.
\]

An incomplete symmetric matrix $A$ is \emph{Robinsonian} (resp., 
\emph{Strong-Robinsonian})
if there is a permutation $\pi:\{1,2,\ldots, n\} \rightarrow 
\{1,2,\ldots, n\}$ such that $A_\pi$, the matrix defined by the entries
$a_{\pi(i),\pi(j)}$, is Robinson (resp., Strong-Robinson). In other 
words, $A$ is Robinsonian (resp., Strong-Robinsonian) if
there is a simultaneous reordering of its rows and columns that results 
in a Robinson (resp., Strong-Robinson) matrix.

\section{Related Work and Our Contributions}\label{sec:relatedwork}
 Recognition of complete Robinsonian matrices has been studied by several
authors. Mirkin et al. in \cite{graphsandgenes84} presented an $O(n^4)$ 
recognition algorithm, where $n\times n$ is the 
size of the matrix. Chepoi et al., using divide and conquer 
techniques, introduced an 
$O(n^3)$ recognition algorithm in \cite{chepoi1997recognition}.
Pr\'ea and Fortin in \cite{prea2014optimal} provided an $O(n^2)$ optimal
recognition algorithm for complete Robinsonian matrices using PQ trees. 

Using the relationship between Robinsonian matrices and unit interval 
graphs presented in \cite{roberts69}, Monique Laurent and Matteo 
Seminaroti  in \cite{laurent2017lex}
introduced a recognition algorithm for Robinsonian matrices that uses 
Lex-BFS, whose time complexity is 
$O(|w|(m+n))$, where $m$ is the number of nonzero entries in the 
matrix,
and $|w|$ is the number of different 
values in the matrix. 
Later in \cite{laurent2017similarity}, the same authors presented a 
recognition algorithm with time complexity $O(n^2+nm\log n )$
that uses similarity first search. 
Again, using the relationship between Robinsonian matrices and unit 
interval graphs, Laurent et al. in \cite{laurent2017structural}
gave a characterization of Robinsonian matrices via forbidden patterns.

The Seriation problem also has been studied as an optimization problem.
Given an  $n\times n$ matrix $D$, \emph{seriation in the presence of 
errors} is to find a Robinsonian matrix $R$ 
that minimizes the error defined as: $\max ||d_{i,j}-r_{i,j}||$ over all 
$i$ and $j$ in $\{1,2,3,\ldots,n\}$. Chepoi et al. in 
\cite{chepoi2009seriation} proved that seriation in the presence of 
errors is NP-Hard. 
Chepoi and Seston in \cite{chepoi2011seriation} gave a factor 16 
approximation algorithm for the same problem. 
Finally, Fortin in \cite{fortin2017robinsonian} surveyed
the challenges for Robinsonian matrix recognition. 

Recognition of $0-1$ incomplete Robinsonian matrices was studied in
\cite{cygan2015sitting} by Cygan et al.,
where it was shown the
NP-Completeness of this problem.
Furthermore, 
the authors concluded that, 
under the assumption of the Exponential Time Hypothesis\footnote{The Exponential Time Hypothesis 
states that there exists a constant $C>0$ such that no algorithm solving the
3-CNF-SAT problem in $O(2^{CN})$ exists, where $N$ denotes the number of variables
in the input formula.},
it is impossible to have a
sub-exponential time algorithm 
that determines if an incomplete symmetric matrix is Robinsonian or not. They also presented an exponential time algorithm for the recognition of incomplete $0-1$ Robinsonian matrices. This algorithm was extended 
in \cite{aracena2019weigtedSCFE} to a recognition 
algorithm of $n\times n$ incomplete Robinsonian matrices with time complexity $O(n\cdot 2^{2n})$ that drops the requirement for the matrix to be $0-1$.

Pardo et al. in \cite{pardo2015embedding} studied an optimization
version for the Seriation problem where the input where $0-1$ incomplete
symmetric matrices. In that work, the 
authors defined 
as an error a violation of one of the inequalities in Definition 
\ref{def:RobinsonIncomplete} and provided heuristics that 
find a reordering of rows and columns attempting to 
minimize the number of errors.
Later Pardo et al. in \cite{pardo2020basic} presented 
a basic variable neighborhood search algorithm for the same problem. 
This algorithm has shown the best experimental results for this problem. 

\paragraph*{Our Contributions}
We consider that our first contribution is the introduction of 
incomplete Strong-Robinson and Strong-Robinsonian matrices.
Then, in Lemma \ref{thm:robnotstrongrob} we show the existence of 
a Robinsonian matrix that is not Strong-Robinsonian. 
In 
Theorem \ref{thm:strongrobNPcomplete}, we show that recognition of 
incomplete Strong-Robinsonian matrices is NP-Complete. Finally, in 
Theorem \ref{thm:parametrizedcomplexity}, we show that recognition of
incomplete Strong-Robinsonian matrices is in XP. Indeed, we 
provide a parameterized algorithm for incomplete Strong-Robinsonian matrix 
recognition that is exponential only in the number of missing entries of
the matrix while polynomial in the size of the matrix and the number of
different values it has.  
\section{Recognition of Strong-Robinsonian Incomplete Matrices}\label{sec:StrongRobRecog}
In this section, we study the recognition problem for incomplete 
Strong-Robinsonian matrices. 
We start introducing some concepts. 
A \emph{completion} of an incomplete symmetric matrix
$A$ is an assignment of values to all the missing entries of $A$.
We say that \emph{a completion of $A$ is Robinsonian} if
the completed matrix is Robinsonian. 
Let $S\subseteq \mathbb{R}$ be a set of real values. 
A completion of $A$ whose new values are 
taken from $S$ is said to be a \emph{completion 
of $A$ in $S$}.

\begin{lemma}\label{lem:StrongRobiifRobcompl}
Let $A$ be an incomplete symmetric matrix. Then $A$ is 
Strong-Robinsonian 
if and only if $A$ has a Robinsonian completion. 
\end{lemma}
\begin{proof}
Let $A$ be an incomplete symmetric matrix. 
If $A$ has a Robinsonian completion, we first complete $A$ according to 
that completion. Then, we write the completion of $A$ in its Robinson 
form. Finally, we delete all the added entries.
The outcome of this process is $A$ written as an incomplete 
Strong-Robinson matrix. 

On the other hand, if $A$ is Strong-Robinsonian, we first write $A$ 
in its Strong-Robinson form. Then, a Robinsonian completion for $A$ is 
constructed as follows. 
For every missing entry in the main diagonal, 
we define $a_{i,i}:=\max w(A)$.
Then, the completion continues filling each diagonal moving away from 
the main diagonal, i. e., increasing the parameter $k=j-i$ for missing 
entries $a_{i,j}$ such that $1\leq i < j \leq n$.
For every missing entry $a_{i,j}$, we define $a_{i,j} := 
\min\{a_{i,j-1},a_{i+1,j}\}$. 
Finally, by construction the completion is Robinson. 
\end{proof}

A direct consequence from the previous lemma is the following lemma.
\begin{lemma}\label{lem:completion}
Let $A$ be an incomplete symmetric matrix, and $w(A)$ be the set of different values in $A$. Then
$A$ has a Robinsonian completion 
if and only if $A$ has a Robinsonian completion in $w(A)$.  
\end{lemma}
\begin{proof}
In one direction, the lemma is direct (if $A$ has a Robinsonian completion in $w(A)$, it has a Robinsonian completion). In the opposite direction, due to Lemma \ref{lem:StrongRobiifRobcompl}, if $A$ has a Robinsonian completion, then $A$ is Robinson. Therefore, we can write it as a Robinson matrix, and then proceed with the completion described in the proof of Lemma \ref{lem:StrongRobiifRobcompl} which provides a Robinsonian completion in $w(A)$.
\end{proof}

In the next lemma we present a Strong-Robinsonian incomplete matrix that is not Robinsonian.

\begin{lemma}\label{thm:robnotstrongrob}
There exists an incomplete Robinsonian symmetric matrix which is not
Strong-Robinsonian.  
\end{lemma}
\begin{proof}
We demonstrate this lemma by giving an incomplete Robinsonian matrix which is not Strong-Robinsonian. 
Consider the following matrix: 
\begin{equation}\label{eq:matrixA}
\begin{pmatrix}
 1 & 1 & 0 & 0 & 0 & 0\\
 1 & 1 & 1 & * & 1 & 0\\
 0 & 1 & 1 & 0 & * & 0\\
 0 & * & 0 & 1 & 1 & 0\\
 0 & 1 & * & 1 & 1 & 1\\
 0 & 0 & 0 & 0 & 1 & 1\\
\end{pmatrix}.
\end{equation}

Matrix \eqref{eq:matrixA} is Robinson, but we shall see that it is not Strong-Robinsonian. By Lemma \ref{lem:StrongRobiifRobcompl}, matrix \eqref{eq:matrixA} is Strong-Robinsonian if and only if it has a Robinsonian completion. By Lemma \ref{lem:completion}, matrix \eqref{eq:matrixA} has a Robinsonian completion if and only if it has a completion with $0$'s and $1$'s. Matrix \eqref{eq:matrixA} has  four possible completions with $0$'s and $1$'s. Two of them generate the same case reordering the columns and rows. Therefore, we analyze three different completions. The three completions are: 
\begin{equation*}\label{eq:matrixAcompl}
\left(
\begin{smallmatrix}
 1 & 1 & 0 & 0 & 0 & 0\\
 1 & 1 & 1 & 0 & 1 & 0\\
 0 & 1 & 1 & 0 & 0 & 0\\
 0 & 0 & 0 & 1 & 1 & 0\\
 0 & 1 & 0 & 1 & 1 & 1\\
 0 & 0 & 0 & 0 & 1 & 1\\
\end{smallmatrix}
\right),
\qquad
\left(
\begin{smallmatrix}
 1 & 1 & 0 & 0 & 0 & 0\\
 1 & 1 & 1 & 1 & 1 & 0\\
 0 & 1 & 1 & 0 & 0 & 0\\
 0 & 1 & 0 & 1 & 1 & 0\\
 0 & 1 & 0 & 1 & 1 & 1\\
 0 & 0 & 0 & 0 & 1 & 1\\
\end{smallmatrix}
\right),
\qquad
\left(
\begin{smallmatrix}
 1 & 1 & 0 & 0 & 0 & 0\\
 1 & 1 & 1 & 1 & 1 & 0\\
 0 & 1 & 1 & 0 & 1 & 0\\
 0 & 1 & 0 & 1 & 1 & 0\\
 0 & 1 & 1 & 1 & 1 & 1\\
 0 & 0 & 0 & 0 & 1 & 1\\
\end{smallmatrix}
\right).
\end{equation*}

Picking the right columns and corresponding rows for each of these matrices, we find the following matrix as a sub matrix:
\begin{equation}\label{eq:matrixB}
\left(\begin{smallmatrix}
1&1&0&0\\
1&1&1&1\\
0&1&1&0\\
0&1&0&1\\
\end{smallmatrix}
\right).
\end{equation}

According to the characterization of matrices of interval and proper interval graphs given by Mertzios in \cite{mertzios2008matrix},
matrix \eqref{eq:matrixB} is not the augmented adjacency matrix 
of a proper interval graph. Due to the result shown in \cite{roberts69} by Roberts, a matrix with $0-1$-entries is Robinsonian if and only if it is the augmented adjacency matrix of a proper interval graph. Therefore, matrix \eqref{eq:matrixB} is not Robinsonian and therefore matrix \eqref{eq:matrixA} is not Robinsonian.
\end{proof}

As a consequence of Lemma \ref{lem:StrongRobiifRobcompl} and Lemma \ref{lem:completion}, we can state the two 
following theorems regarding the complexity of incomplete 
Strong-Robinsonian matrix recognition.  

\begin{theorem}\label{thm:strongrobNPcomplete}
Strong-Robinsonian matrix recognition is NP-Complete. 
\end{theorem}
\begin{proof}
We show the NP-Completeness of this recognition problem
via a particular case. Indeed, we will show that Strong-Robinsonian
$0-1$ matrix recognition is NP-Complete. 

From Lemma \ref{lem:StrongRobiifRobcompl} and Lemma \ref{lem:completion}
we conclude that a $0-1$ incomplete symmetric matrix $A$ is 
Strong-Robinsonian if and only if $A$ has a Robinsonian completion with 
$0$'s and $1$'s. On the other hand, in \cite{cygan2015sitting} it was 
shown that a complete symmetric $0-1$ matrix 
is Robinsonian if and only if it is the augmented adjacency matrix of a
unit interval graph. 
The problem is in NP since any $0-1$ completion can be tested to be Robinsonian in polynomial time.  
Moreover, using the NP-Completeness of the 
matrix sandwich problem for adjacency matrices of unit interval graphs shown in \cite{GOLUMBIC1994251} and \cite{GOLUMBIC1998239}, we 
conclude that Strong-Robinsonian matrix recognition is NP-Complete.
\end{proof}

Finally, we show that Strong-Robinsonian matrix recognition is fixed-parameter tractable. 
\begin{theorem}\label{thm:parametrizedcomplexity}
Let $A$ be an incomplete symmetric matrix with $b$ missing entries and $|w(A)|$ different 
values. Then, it is possible to
decide if $A$ is Strong-Robinsonian in time $O(|w(A)|^bn^2)$. 
\end{theorem}
\begin{proof}
Let $A$ be an incomplete symmetric matrix with $b$ missing entries and $|w(A)|$ different 
values.
There exist $|w(A)|^b$ different completions of $A$ with values in 
$w(A)$.
Therefore, an exhaustive search over all the completions of $A$ with
values in $w(A)$, and testing for each of them the Robinsonian 
property (for complete matrices)
can be done in time $O(|w(A)|^bn^2)$.
\end{proof}

\section{Concluding Remarks}\label{sec:concludingremarks}
Robinsonian matrices are important in different contexts. 
Recognition of Robinsonian matrices can be done efficiently when the matrix has no missing entry. When entries of the matrix are missing, 
Robinsonian matrix recognition can be done in $O(n\cdot 2^n)$ time. Unfortunately, it has been shown that there is no sub-exponential time algorithm for the Robinsonian (incomplete) matrix recognition problem.

In this document, we proved that a subset of the set of incomplete Robinsonian matrices, the Strong-Robinsonian matrices, can be recognized in polynomial time with respect to the size of the matrix when the number of missing entries is a constant parameter. Indeed, we presented an algorithm that recognizes incomplete Strong-Robinsonian matrices in time $O(|w|^bn^2)$, where $|w|$ denotes the number of different values in the matrix and $b$ denotes the number of missing entries.  


\end{document}